\documentclass[apj]{emulateapj}
\usepackage{amsmath}
\usepackage{natbib}
\usepackage{graphics}
\usepackage{appendix}
\usepackage{color}
\usepackage{relsize}
\shorttitle{Ground-based observations of 55Cnc e}
\shortauthors{De Mooij et al.}

\begin{document}

   \title{Ground-Based Transit Observations of the Super-Earth 55 Cnc e}
   \author{E.J.W. de Mooij \altaffilmark{1},
           M. L\'opez-Morales\altaffilmark{2},
           R. karjalainen \altaffilmark{3},
           M. Hrudkova\altaffilmark{3},
           and           
           R. Jayawardhana\altaffilmark{4}}
\email{demooij@astro.utoronto.ca}
\altaffiltext{1}{Astronomy and Astrophysics,  University of Toronto, Toronto, Canada}
\altaffiltext{2}{Harvard-Smithsonian Center for Astrophysics, 60 Garden St., Cambridge, MA  USA}
\altaffiltext{3}{Isaac Newton Group of Telescopes, La Palma, Spain}
\altaffiltext{4}{Physics \& Astronomy, York University, Toronto, Canada}

\begin{abstract}
We report the first ground-based detections of the shallow transit of the super-Earth 
exoplanet 55 Cnc e using a 2-meter-class telescope. Using differential spectrophotometry, 
we observed one transit in 2013 and another in 2014, with average spectral resolutions 
of $\sim$700 and $\sim$250, spanning the Johnson BVR photometric bands. We find a 
white-light planet-to-star radius ratio of 
0.0190$_{-0.0027}^{+0.0023}$ from the 2013 observations and 0.0200$_{-0.0018}^{+0.0017}$
from the 2014 observations. The two datasets combined results in a radius ratio 
of 0.0198$_{-0.0014}^{+0.0013}$. These values are all 
in agreement with previous space-based results. Scintillation noise in the data 
prevents us from placing strong constraints on the presence of an extended
hydrogen-rich atmosphere. Nevertheless, our detections of 55 Cnc e in transit 
demonstrate that moderate-size telescopes on the ground will be capable of routine 
follow-up observations of super-Earth candidates discovered by the Transiting 
Exoplanet Survey Satellite (TESS) around bright stars. We expect it will be also 
possible to place constraints on the atmospheric characteristics of those planets 
by devising observational strategies to minimize scintillation noise.
\end{abstract}


\section{Introduction}

Confirmation, follow-up, and atmospheric characterization of a large number of super-Earths and smaller planets will be 
among the main challenges facing exoplanet researchers in the next decade, especially once missions such as TESS 
\citep{ricker2014} start identifying numerous candidates around bright, nearby stars. 

Current follow-up plans include the confirmation of those planet candidates using well tested ultra-high precision radial velocity instruments --- like HARPS and HARPS-N \citep[see e.g.][]{lopezmoralesetal2014}, and future instruments like ESPRESSO on the VLT \citep{pepeetal2010}, G-CLEF on the GMT \citep{szentgyorgyietal2012}, and CODEX on the E-ELT \citep{pasquinietal2010} --- and the atmospheric characterization of the most interesting confirmed planets using future ground-based (e.g. GMT, TMT, and E-ELT),  and space-based (e.g. JWST) facilities. However, not much attention is being paid to the potential role of smaller ground-based telescopes in the confirmation and follow-up of those planets. While there is some work underway to build small arrays of robotic telescopes for radial velocity and photometric follow-up, e.g. Project MINERVA\footnote{http://exolab.caltech.edu/research/minerva.html}, the full capacity of moderate-size existing telescopes and instruments for this task has not been explored yet.

In this Letter, we report the results of our attempts to detect the shallow transit and the atmosphere of a super-Earth around a bright nearby star with a 2-meter class telescope and current instrumentation on the ground. Our target is 55 Cnc e, the super-Earth with $M_{\rm p}$ = 8.3 $\pm$ 0.4 $M_{\rm Earth}$; $R_{\rm p}$ = 1.94 $\pm$ 0.08 $R_{\rm Earth}$, found to transit by \citet{winnetal2011} and \citet{demoryetal2011} after \citet{dawsonetal2010} provided a revised period of 0.74 days, shorter than the originally reported 2.8-day period derived by \citet{mcarthuretal2004}. The transit depth of 55 Cnc e is only $\sim$ 0.4 mmag, and it orbits the brightest star known to harbor a transiting planet, 55 Cnc (V = 5.95). Therefore, 55 Cnc is, of all currently known planet hosts, the one with characteristics most similar to the targets that TESS will find. 55 Cnc is a 0.9 $M_{\sun}$ star \citep{vonbraunetal2011} and it harbors four other planets with masses between  0.14 $M_{\rm Jup}$ and 3.8 $M_{\rm Jup}$, and orbital periods between $\sim$ 14.6 days and  $\sim$ 4872 days \citep{nelsonetal2014}.

\section{Observations}\label{sec:obs}

We observed two transits of 55 Cnc e with the ALFOSC instrument on the 2.5-meter Nordic Optical Telescope (NOT) at the Observatorio del Roque de los Muchachos in La Palma, Spain, as part of the NOT Fast-Track Service program.

\subsection{Transit observations on 28 March 2013}
 The observations started at 20:54 on March 28 2013 UT (hereafter Night 1), and lasted four hours, during which we obtained 602 frames. The exposure time was 10 seconds, and we used the 400kpix/sec readout mode to reduce overheads, providing an average cadence of 24 seconds. We used the nearby M-star BO Cnc as a reference to correct for instrumental and atmospheric effects. 
 
 Both stars were observed simultaneously in differential spectrophotometry mode with the \#7 grism, which provides a wavelength coverage from 3800$\AA$ to 6800$\AA$. To prevent slit losses, we opted to perform these observations without a slit. We used the 550\_275 broadband filter to reduce sky background at the expense of reducing spectral coverage to 4100$\AA$ -- 6800$\AA$. Note that the transmission profile of the  550\_275 filter is not flat, but shows variations of up to 20\% over 100$\AA$ wavelength scales\footnote{http://www.not.iac.es/instruments/filters/curves/png/92.png}. Since the observations were slitless, the spectral resolution was determined by the seeing. The median seeing was $\sim$1 arcsec (6 pixels) during the night, which resulted in a resolution of $\sim$700.

During the observations the target passed close to the zenith, requiring a very rapid movement by the derotator. This caused the position angle to vary by more than 0.2$^{\circ}$ for $\sim$9 minutes. The offset in the rotator angle resulted in a blurring of the PSF both spatially and spectrally, as well as an offset in the spatial and spectral directions. We therefore exclude all frames where the offset in position angle was larger than 0.2$^{\circ}$; in total 24 frames were removed due to this. Also, many of the frames reached peak counts above the linearity limit of the detector ($\sim$300,000 ADU).

\subsection{Transit observations of February 9, 2014}
A second transit of 55 Cnc e was observed between 01:05 and 05:00 February 9, 2014 UT (hereafter Night 2), resulting in a  total of 556 frames. Exposure time, readout mode, and cadence were the same as in Night 1.

In contrast to Night 1, the Night 2 observations utilized  a 40$\arcsec$ slit and no filter. The slit width was chosen to prevent slit losses due to seeing fluctuations and guiding errors, while significantly reducing the sky background. As in Night 1, we used the \#7 grism, which provides spectral coverage between 3800$\AA$ and 6800$\AA$. All the observations were obtained after the target's meridian crossing, so we experienced no derotator problems.

The telescope was significantly defocused in order to keep the flux of the target and reference stars within the detector's  linear regime. This resulted in a spatial FWHM of $\sim$14-15 pixels, corresponding to $\sim$2.6-2.9'', given the 0.19''/pixel platescale of ALFOSC. This yields an average resolution of $\sim$250. The first 26 frames were not defocused and were discarded.

\section{Data reduction and analysis}\label{sect:dr}
\begin{figure*}
\centering
\includegraphics[width=18cm]{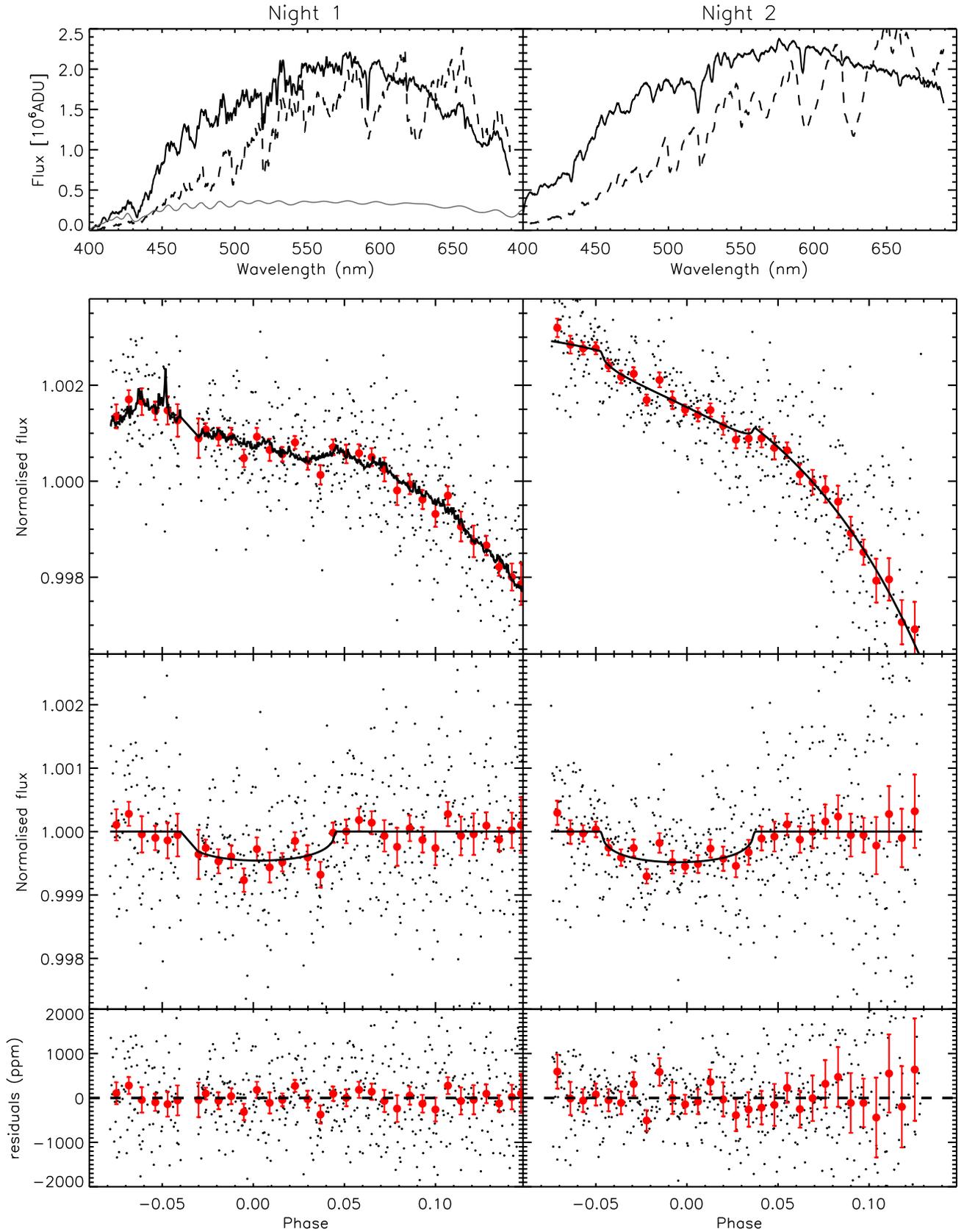}
\caption{Example spectra (top panels)  and the white-light curves for Night 1 (left) and Night 2 (right). The second row shows the differential light curves with their best-fit models overplotted. The third row of panels shows the light curves after removing the baseline trends. The residuals are shown in the bottom panels. Small dots represent the unbinned data. Big red dots show the data binned in $\sim$7.5 min intervals ($\sim$0.007 in phase). In the top panel the filtercurve (light grey line) is plotted together with the spectra of the target (solid lines) and reference (dashed lines) stars.}
\label{fig:white_lc_indiv}
\end{figure*}
\begin{figure*}
\centering
\includegraphics[width=18cm]{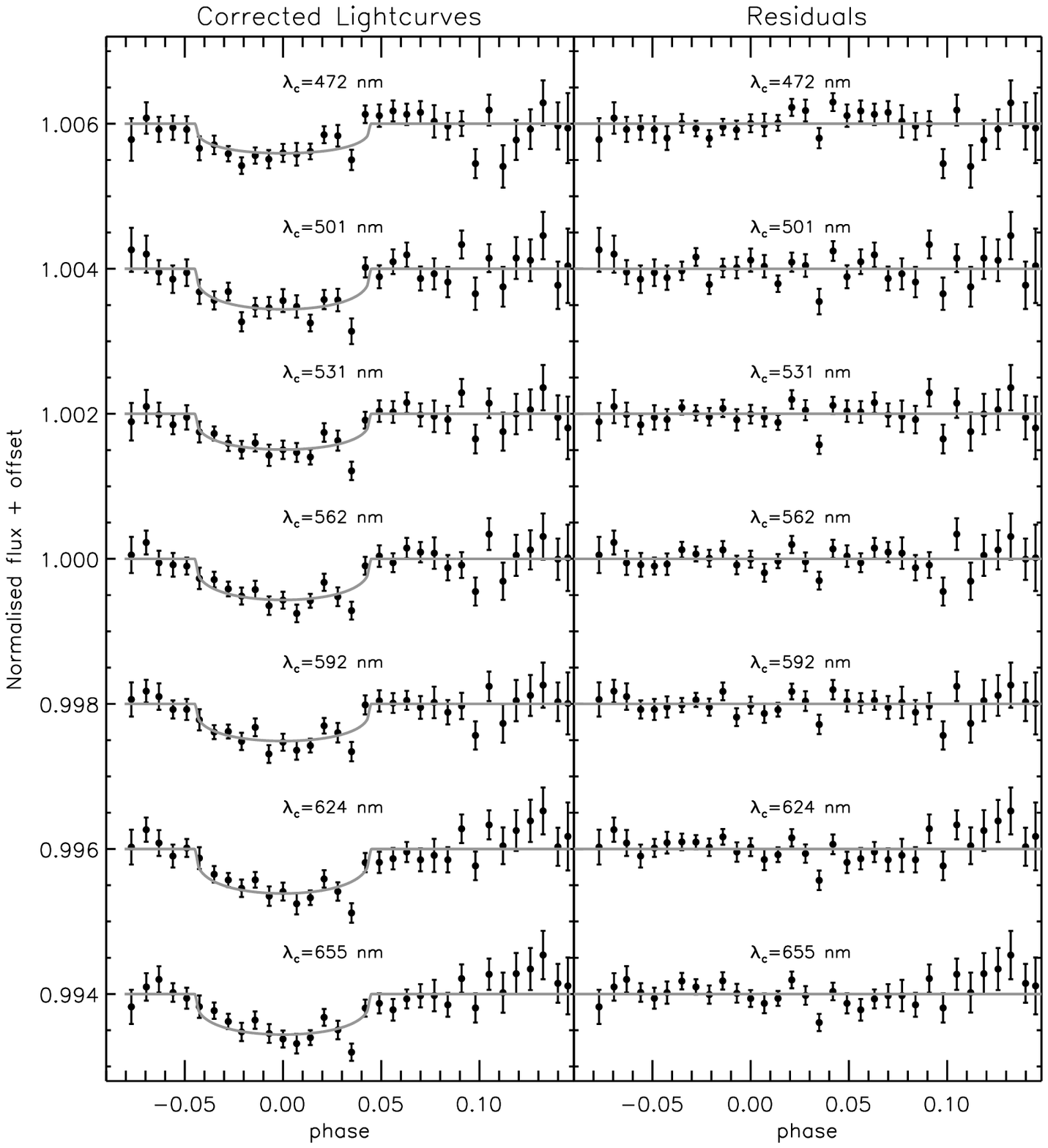}
\caption{Light curves of each of the individual 200 pixel ($\sim$30 nm) bins. The left panel shows the light curves for both nights combined, after removing the baseline trends. The right panel shows the residuals of the fits. The dots correspond to the data binned by $\sim$7.5 minutes  ($\sim$0.007 in phase). The lines show the best transit models and the labels on top of each light curve show the central wavelength of each bin. An offset has been applied between the light curves for clarity.}
\label{fig:color_lc}
\end{figure*}

\subsection{Data reduction}

We reduced the data from both nights using a custom set of IDL routines. We first trimmed the overscan and bias-subtracted the frames. For Night 1 we applied a non-linearity correction to account for the effect of count levels in the non-linearity regime. The linearity correction was based on calibration data obtained by the NOT staff for the 400 kpix/sec read-out mode. Subsequently, we flatfielded the data and removed the sky background using a slightly modified version of the method proposed by \cite{kelson2003}. For each star the sky outside the stellar PSF was fit with a spline function using BSPLINE from IDL Utils\footnote{www.sdss3.org/dr10/software/idlutils}; any residual gradient in the background was accounted for by fitting a low-order two dimensional polynomial surface. Finally, we extracted the spectra of each star by summing up the flux in the spatial direction in a region of $\pm$110 pixels and $\pm$80 pixels from the center of the trace for Night 1 and 2, respectively. We verified that the aperture choice does not alter the results as long as it encompasses most of the PSF. Example extracted spectra for each object, in each night,  are shown in Figure~\ref{fig:white_lc_indiv}. 

As a final step before generating the light-curves, we re-sampled the spectra onto a common wavelength grid. This is necessary because no atmospheric dispersion corrector was used for the observations and, to observe 55 Cnc and the reference star simultaneously, the slit was not aligned with the paralactic angle. Furthermore, residual guiding errors and telescope drift in the spectral direction, also result in wavelength offsets.

Light curves were generated by binning the extracted spectra in wavelength. For the white-light curves we integrated over 1400 pixels in wavelength, which corresponds to $\sim$214nm. The narrower color-channel light curves were generated using 200 pixel bins ($\sim$30nm). The white-light curves for both nights are shown in Figure ~\ref{fig:white_lc_indiv}. The light curves for the different color channels are shown in Figure~\ref{fig:color_lc}.

\subsection{Light curve fitting}
\begin{figure*}
\centering
\includegraphics[width=18cm]{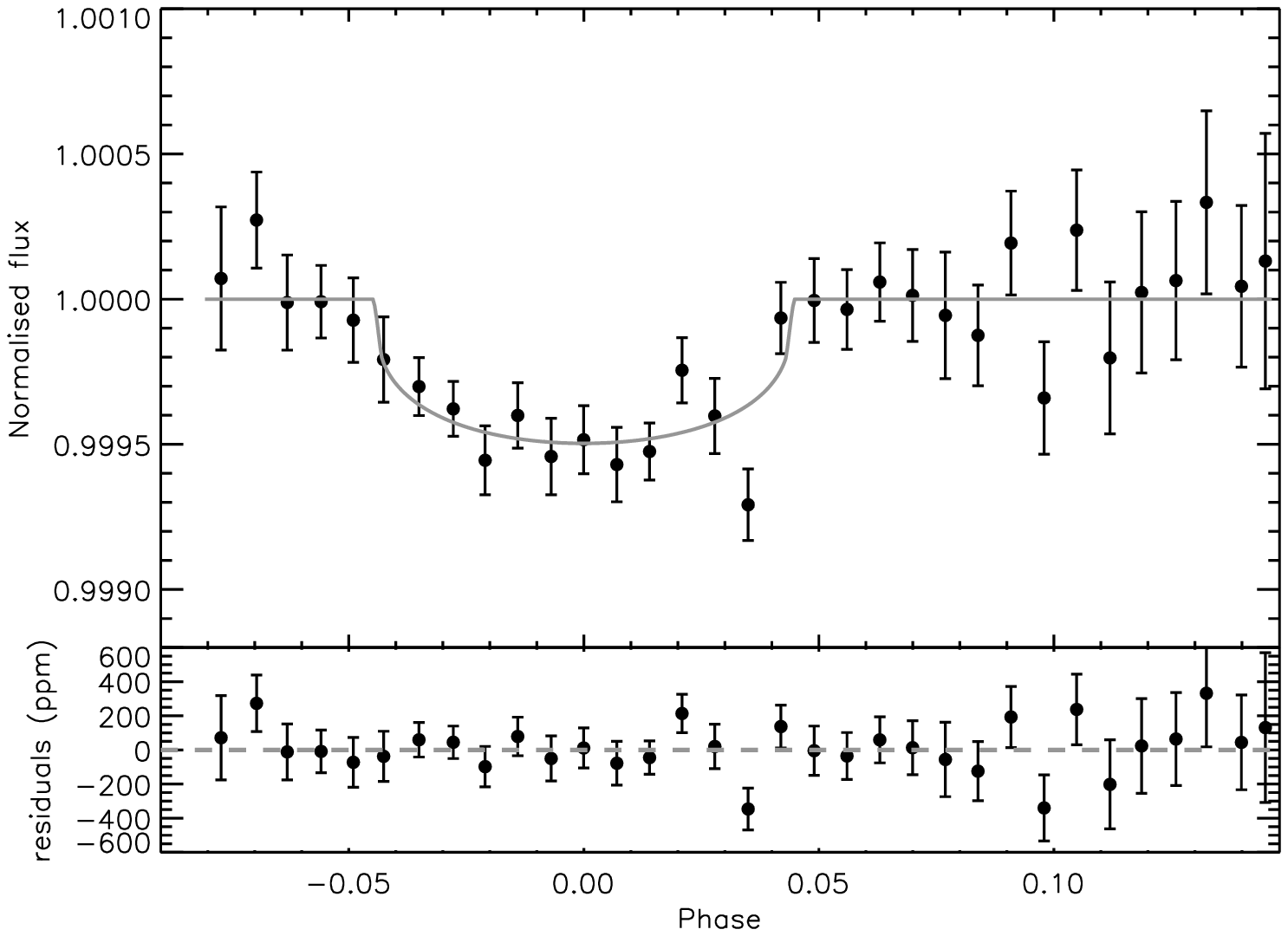}
\caption{The combined and phase-folded light-curve for both nights after correction for systematic effects (top panel). The residuals are shown in the bottom panel. The dots show the data binned by 0.007 in phase ($\sim$7.5 min).}
\label{fig:white_lc_comb}
\end{figure*}

As revealed in the second row of Fig.~\ref{fig:white_lc_indiv}, the differential light curves from both nights show significant trends. We fit those trends together with the parameters of the transit. The parameters we used to model the trends were the spectral centroid position, airmass, the FWHM in the spatial direction (seeing), and residual sky background.  For the transit, we fit the planet-to-star radius ratio and the mid-transit time, modelled as an offset in phase. 

The parameter values for the best baseline models were obtained using a Bayesian Inference Criterium (BIC) analysis \citep[e.g.][]{liddle2007}, where the best solution is given by the lowest BIC value. To model the transits we used the \citet{mandelandagol2002} code, with limb darkening coefficients computed from the \citet{claret2000} tables, for a star of properties similar to 55 Cnc \citep{vonbraunetal2011}, and assuming an effective wavelength passbands similar to V-band.  Observational noise dominates the errors for these observations, so the limb darkening coefficients cannot be reliably constrained. Also, because of the relatively high noise levels, we needed to fix the planet's relative separation a/R$_*$ and impact parameter {\rm b} to the values from \citet{dragomiretal2014}.

In addition to fitting the nights individually, we also performed joined fits for the two nights.  We fitted two different models to the combined light curves. In the first model we allowed the mid-transit time for each night to vary independently, while in the second model the phase offset of mid-transit, although allowed to vary, was the same for both nights. The light curves for the color channels were fit jointly for the two nights, with the time of mid-transit fixed to that obtained from the fit to the white-light curves.
The fitting was done using a Markov-Chain  Monte-Carlo procedure, for which we ran five chains of 200,000 steps each, discarding the first 20,000 steps to prevent initial parameter biases.  The five chains were then merged, after verifying that they were well mixed \citep{gelmanandrubin1992}. The parameters perturbed for all channels and runs are R$_p$/R$_*$ and the coefficients of the baseline. For the white lightcurves we also perturbed T$_0$, and included a/R$_*$ and b as priors. The best fit parameters were determined using the 50\% value of the distribution, and the uncertainties were obtained from the 16-84\% confidence interval. The best fit models are overplotted in Fig.~\ref{fig:white_lc_indiv} for the individual white curves, in Fig.~\ref{fig:color_lc} for the combined color channel curves, and in Figure \ref{fig:white_lc_comb} for the combined white curve. The corresponding parameter values for each fit are listed in Table 1.

\section{Results}\label{sect:discus}

\begin{table}
\caption{Best fit planet-to-star radius ratio  (R$_p$/R$_*$) and T0 offset (in terms of orbital phase, $\Delta\phi_i$) values for white and color channel, individual and combined light curves. }
\centering
\begin{tabular}{ccccc}
\hline
Band    & Wavelength    &       R$_p$/R$_*$          &         $\Delta\phi_1$       &         $\Delta\phi_2$       \\
        &    (nm)       &                            &                             &                              \\
\hline
\multicolumn{5}{c}{\bf{Individual lightcurves}}\\
 White   &  457-671   &  0.0190$_{-0.0027}^{+0.0023}$ &   -0.0026$_{-0.0065}^{+0.0052}$  &                     $-$      \\
 White   &  457-671   &  0.0200$_{-0.0018}^{+0.0017}$ &           $-$                &    0.0048$_{-0.0032}^{+0.0025}$ \\
         &            &                            &           \\
\hline
\multicolumn{5}{c}{\bf{Combined, independent T0 offsets} }  \\
 White   &  457-671   &  0.0198$_{-0.0014}^{+0.0013}$ &   -0.0020$_{-0.0053}^{+0.0038}$  &    0.0044$_{-0.0026}^{+0.0027}$  \\
   C1    &  459-488   &  0.0180$_{-0.0017}^{+0.0016}$ &   -0.0020\textsuperscript{a} &    0.0044\textsuperscript{a} \\
   C2    &  488-517   &  0.0211$_{-0.0015}^{+0.0014}$ &   -0.0020\textsuperscript{a} &    0.0044\textsuperscript{a} \\
   C3    &  517-547   &  0.0198$_{-0.0015}^{+0.0014}$ &   -0.0020\textsuperscript{a} &    0.0044\textsuperscript{a} \\
   C4    &  547-578   &  0.0212$_{-0.0013}^{+0.0013}$ &   -0.0020\textsuperscript{a} &    0.0044\textsuperscript{a} \\
   C5    &  578-609   &  0.0201$_{-0.0014}^{+0.0013}$ &   -0.0020\textsuperscript{a} &    0.0044\textsuperscript{a} \\
   C6    &  609-640   &  0.0221$_{-0.0013}^{+0.0012}$ &   -0.0020\textsuperscript{a} &    0.0044\textsuperscript{a} \\
   C7    &  640-672   &  0.0211$_{-0.0013}^{+0.0012}$ &   -0.0020\textsuperscript{a} &    0.0044\textsuperscript{a} \\
         &            &                            &           \\
\hline
\multicolumn{5}{c}{\bf{Combined, common T0 offset}}\\
 White   &  457-671   &  0.0195$_{-0.0014}^{+0.0013}$ &    0.0017$_{-0.0036}^{+0.0027}$  &   0.0017$_{-0.0036}^{+0.0027}$  \\
   C1    &  459-488   &  0.0175$_{-0.0018}^{+0.0016}$ &    0.0017\textsuperscript{a} &    0.0017\textsuperscript{a} \\
   C2    &  488-517   &  0.0208$_{-0.0015}^{+0.0014}$ &    0.0017\textsuperscript{a} &    0.0017\textsuperscript{a} \\
   C3    &  517-547   &  0.0198$_{-0.0015}^{+0.0014}$ &    0.0017\textsuperscript{a} &    0.0017\textsuperscript{a} \\
   C4    &  547-578   &  0.0208$_{-0.0014}^{+0.0013}$ &    0.0017\textsuperscript{a} &    0.0017\textsuperscript{a} \\
   C5    &  578-609   &  0.0199$_{-0.0014}^{+0.0013}$ &    0.0017\textsuperscript{a} &    0.0017\textsuperscript{a} \\
   C6    &  609-640   &  0.0220$_{-0.0013}^{+0.0012}$ &    0.0017\textsuperscript{a} &    0.0017\textsuperscript{a} \\
   C7    &  640-672   &  0.0210$_{-0.0013}^{+0.0012}$ &    0.0017\textsuperscript{a} &    0.0017\textsuperscript{a} \\
\hline
\end{tabular}

\textsuperscript{a} Phase offset fixed to value from white lightcurve.
\label{tab:rprs}
\end{table}

\begin{figure*}
\centering
\includegraphics[width=17.7cm]{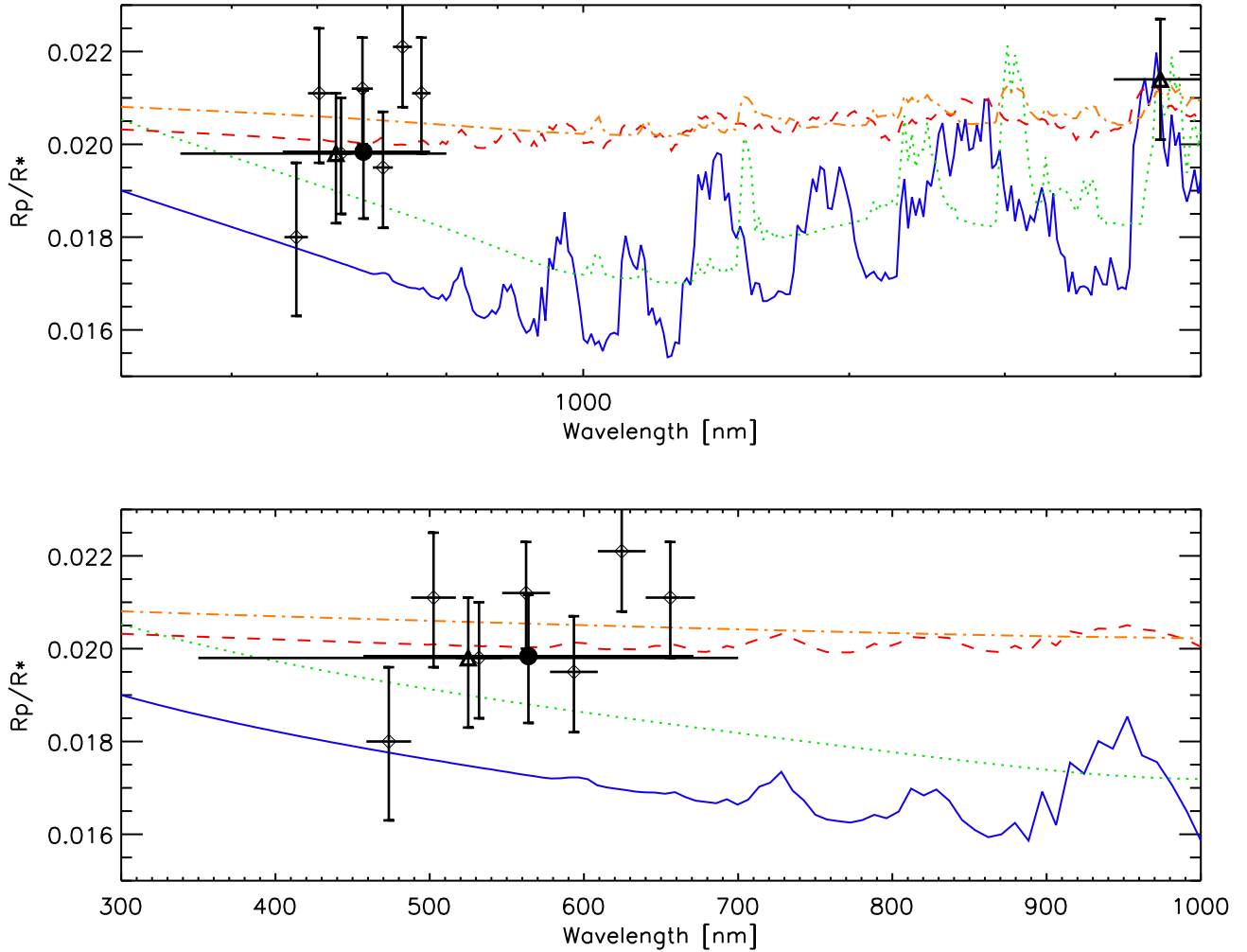}
\caption{Comparison of our $R_p / R_*$ ratios in Table 1 (open diamonds: 30nm bins \& filled circle: white light curve bin) to $H_2$-dominated and non $H_2$-dominated atmosphere models for 55 Cnc e, for carbon-rich and oxygen-rich atmospheres \citep{hu2014}. The open triangles correspond to the $R_p / R_*$ ratios measured with $MOST$ and $Spitzer$ \citep{gillonetal2012}. The models shown are for a hydrogen dominated atmopshere with C/O=0.5 (solid blue line) and C/O=2 (dotted green line) as well as a hydrogen poor atmosphere with C/O=0.5 (dashed red line) and C/O=2 (orange dash-dotted line). }
\label{fig:transmission_spec}
\end{figure*}

We measure a  white-light radius ratio for 55 Cnc e of 0.0190$_{-0.0027}^{+0.0023}$ on Night 1 and 0.0200$_{-0.0018}^{+0.0017}$ on Night 2. For the combined dataset, we obtain a white-light radius ratio of 0.0198$_{-0.0014}^{+0.0013}$, and find that a model with a separate T$_0$ offset for each night is slightly favored over a model where the T$_0$ offsets between nights are fixed. However, both T$_0$ offsets are consisted with zero, revealing no noticeable transit timing variations between the two epochs.
When we force the T$_0$ offset to be identical for both nights (Model 2 in Table~\ref{tab:rprs}), we again find that the phase offset is small, and that the planet-to-star radius-ratio is unaltered from the free phase offset model that provided the lowest BIC value.

Our white-light radius ratios are fully consistent with both the {\it MOST} satellite and {\it Spitzer} measurements from~\citet{gillonetal2012} and \citet{dragomiretal2014}. When binning the residuals of the white light curves for each of the nights the RMS decreases steeper than 1/sqrt(N), with N the number of points in the bin, indicating that any contribution from red-noise is small or hidden by the much larger scintillation effects (see below).

The color channel radius ratios summarised in Table~\ref{tab:rprs} are also plotted in Fig.~\ref{fig:transmission_spec}, and compared to the transmission spectra models for 55 Cnc e generated by \citet{hu2014}. The models are for a $H_2$-dominated atmosphere ($X_{H}$ = 0.99) and a non-$H_2$-dominated atmosphere ($X_{H}$ = 0.5), with two different carbon-to-oxygen ratios: a carbon-rich (${\rm C / O}$ = 2.0), and an oxygen-rich atmosphere (${\rm C / O}$ = 0.5).

Our transit depths agree, on average, with the transit depth measured by MOST \citep{gillonetal2012}. However, the errorbars introduced by scintillation limit the capability of our data to place tight constraints on the atmospheric composition of 55 Cnc e, and it will be necessary to improve the errorbars in each bin by at least a factor of two to be able to differentiate between the $H_2$-dominated and non-$H_2$-dominated models at 3-$\sigma$ or larger confidence levels.

As revealed in the white-light curve from Night 2 (right side, bottom panels in Fig.~1), where the observations span a wider range of airmass than in Night 1, the RMS of the light-curve increases rapidly towards the end of the observations. We attribute this increase in the noise to an increase in scintillation at higher airmass. Using the scintillation equation in \citet{hartmanetal2005}, and values of the observatory height above sea level (2383 m) and the telescope diameter (256 cm) for NOT, the exposure time in seconds, an average airmass $z \sim 1.5$, and an standard exponent for the airmass term of $\alpha$ = 1.75, we find the average scintillation noise level is about 1.0 mmag per exposure, while the average differential poisson noise is about 0.05 mmag for the white-light curves, and 0.1--0.2 mmag for the colour channels light curves. Therefore, finding a way to reduce the scintillation noise levels will have important implications for ground-based atmospheric studies of these planets.

\section{Discussion and Summary}\label{sect:summ}

We have presented the first ground-based detection of the shallow transit of the super-Earth 55 Cnc e. The detection was achieved with a 2m-class telescope, and a standard low resolution spectrograph in slitless and wide-slit, differential spectroscopy mode, to allow for the collection of a large number of photons without saturating.

The precision of our transit detection is comparable to that of the previously reported transit detections of this planet from space, with {\it MOST} and {\it Spitzer}, which 
reveals the great potential to do this type of science from the ground, specially in the upcoming TESS era, where the number of small planet candidates around stars in need of follow-up will largely exceed the capacity of space-based and large ground-based instruments. The fact that such follow-up can be carried out with existing instruments on moderate-size telescopes also constitutes a great use of those facilities and a cost effective option.

This alternative is, however, not free of limitations, most of which can be  overcome. In the case of bright stars, scintillation noise is the dominant limitation for small telescopes, but scintillation can be significantly reduced with some intelligent planning.  With the current instrumentation, the main way to do this is to use longer exposure times and thereby reduce the overheads. However, longer exposure times will likely result in saturation of the detector, which can be mitigated by either defocusing the telescope (resulting in a lower spectral resolution, and possible slitlosses), using a higher resolution grating (which will decrease the wavelength coverage), or using a neutral density filter to reduce the stellar flux. The latter option will be offered soon at the William Hershel Telescope. Another possible solution is to increase default gain levels of standard CCDs.

Another constraint is the limited number of bright stars that would fall in the field of view of the instruments to be used as comparison objects to minimize the effect of the Earth's atmosphere. However, this problem can be overcome, at least in the case of color channel spectroscopy. In our observations we find that the scintillation noise between two wavelength channels is highly correlated, and using one wavelength channel as a reference for others significantly reduces the point-to-point scatter, although for our data residual systematic effects dominate the noise level. Another possible solution is to develop a system of pickup mirrors that would redirect the light of comparison stars in the telescope's focal plane in to the detectors.

Our result highlights the potential of using ground-based telescope to perform follow-up of bright transiting planet systems, such as those that will be found by the TESS mission. For future observations of bright super-Earth systems we recommend that large telescopes are used in combination with both a neutral density filter and a (small) defocus in order to allow for more efficient observations and to reduce the impact from scintillation noise.

\subsection*{Acknowledgements}
Based on observations made with the Nordic Optical Telescope, operated by the Nordic Optical Telescope Scientific Association at the Observatorio del Roque de los Muchachos, La Palma, Spain, of the Instituto de Astrofisica de Canarias. 
The data presented here were obtained with ALFOSC, which is provided by the Instituto de Astrofisica de Andalucia (IAA) under a joint agreement with the University of Copenhagen and NOTSA. 
EdM is supported in part by an Ontario Postdoctoral Fellowship. This work is supported by grants to RJ from the Natural Sciences and Engineering Research Council of Canada. 
MLM acknowledges support from a grant from the John Templeton Foundation. The opinions expressed in this publication are those of the authors and do not necessarily reflect the views of the John Templeton Foundation.
 We are grateful to the staff of the Nordic Optical Telescope for their help with executing these observations. We also thank Renyu Hu for sharing his 55 Cnc e atmospheric models with us. We thank the anonymous referee for constructive comments.

\end{document}